\documentstyle[12pt]{article}
\newtheorem{definition}{Definition}
\newtheorem{proposition}{Proposition}
\newtheorem{properties}{\sc Property}
\newtheorem{corollary}{Corollary}
\newtheorem{exemple}{\sc Exemple}

\def\be{\begin{equation}}
\def\ee{\end{equation}}
\def\bea{\begin{eqnarray}}
\def\eea{\end{eqnarray}}

\def\bdf{\begin{definition}}
\def\edf{\end{definition}}
\def\bpr{\begin{properties}}
\def\epr{\end{properties}}
\def\bpt{\begin{proposition}}
\def\ept{\end{proposition}}
\def\bcll{\begin{corollary}}
\def\ecll{\end{corollary}}
\def\bex{\begin{exemple}}
\def\eex{\end{exemple}}

\begin{document}

\renewcommand{\thefootnote}{\fnsymbol{footnote}}
\def\scf{\setcounter{footnote}}


\title{Comment about the Letter entitled ``Scalar fields as dark
 matter in spiral galaxies" }

\author{Daniel Sudarsky \\ \small {\it Instituto de Ciencias Nucleares}
\\ \small
{\it Universidad Nacional Aut\'onoma de M\'exico} \\ \small {\it
Apdo. Postal 70-543 M\'exico D.F. 04510, M\'exico} }
\maketitle

\bigskip

The objective of  this brief comment is to
point out several problems associated  with the  general framework underpining
 this  paper and  with the
application in this work   in particular. For one, it is easily proven that
there are no
non-singular  asymptotically flat,  static solutions to the
coupled Einstein Scalar Field theory. Thus, the solution(s) presented
in \cite{Ton}, ought to be singular, and
therefore can not be considered as corresponding to a real physical galaxy,
in whose
center we expect to have either a
black hole or just a region  with high density ordinary matter.
One could have hoped that one of the situations above would allow one to remove
the singularity. Unfortunately this  will not work with either
of these two alternatives because, in the black hole case, the no hair theorems
 would rule out the static solution  from
the start, and in the ordinary matter case,  the very stringent
experimental bounds for the coupling of such
a scalar field with ordinary matter would disallow the consideration of
ordinary matter as the
``source" of the field.   Moreover if one takes the view that  ordinary
matter at the galactic
center is responsible for the nontrivial configuration of the scalar field,
one needs
to explain why for example a similar phenomena
doesn't happened, say, at the solar system scale.

On the other hand if we take a close look  at the analysis of the motion of
test
particles in \cite{Ton}   we  find a  very serious flaw: in
arriving at eq. 18
 which is supposed to give the condition for circular orbits the authors
 just set $\dot\rho =0$ in the  ``energy
conservation equation" without verifying that the point corresponds to a
 minimum of the effective potential.
 This  would correspond to finding the  turning  points  of the various
non-circular orbits  rather than  the condition
for circular orbits which is what is needed. Moreover, one can go directly to
the final expression for the metric, eq.(21) of \cite{Ton}
, and see explicitly that there are in general no circular orbits as
 the effective potential for the motion on the
equatorial plane is $ V_{eff}(r) = (f_0/r_0) ( (r-a)^2 + K^2 +f_0 r_0 B^2)
$ which has
 no minima other than $r=a$, a point
where the coordinates are not longer valid. Thus there are no "rotation
curves"
associated with the spacetime described
in this  work. Finally even if there were such orbits in the spacetime
associated with the scalar field configuration
and they yielded the expression  in eq. 20 for the velocity of rotation, one
 could not justify the procedure used by the
authors to take into account the effects of the visible matter. Namely to
take the resulting velocity to be
$v_T = \sqrt{V_{DM}^2 + V_L^2}$ where $V_L$ represents somehow the ``circular
velocity of a star
which would be associated with the  luminous matter". There is absolutely no
justification for
this  pythagorean addition of velocities since they are not orthogonal nor do
they represent
statistically uncorrelated averages.

We must therefore conclude that our colleagues have unfortunately  been led
astray
by the enthusiasm in their mistaken  belief of having found a solution to
this most intriguing
problem.

\section*{Acknowledgments} This work was in part supported by
DGAPA-UNAM grant No IN121298 and  by CONACyT
grant 32272-E.

\end{document}